\documentclass[twocolumn]{aastex701}

\begin{document}

\title{Blowing star formation away in AGN Hosts (BAH) - VI. Warm Molecular Outflows Driven by the Low-Luminosity AGN in NGC~3884}

\author[0000-0003-3667-9716]{José Henrique Costa-Souza}
\affiliation{Departamento de F\'isica, CCNE, Universidade Federal de Santa Maria, Av. Roraima 1000, 97105-900,  Santa Maria, RS, Brazil}
\email[show]{henrique.souza@acad.ufsm.br}  

\author[0000-0003-0483-3723]{Rogemar A. Riffel}
\affiliation{Departamento de F\'isica, CCNE, Universidade Federal de Santa Maria, Av. Roraima 1000, 97105-900,  Santa Maria, RS, Brazil}
\email[show]{rogemar@ufsm.br}  

 \author[0000-0001-6100-6869]{Nadia L. Zakamska}
 \affiliation{Department of Physics \& Astronomy, Johns Hopkins University, Bloomberg Center, 3400 N. Charles St, Baltimore, MD 21218, USA}
 \email[noshow]{zakamska@jhu.edu}  

 \author[0000-0002-6570-9446]{Marina Bianchin}
 \affiliation{Instituto de Astrof\' isica de Canarias, Calle V\'ia L\'actea, s/n, E-38205, La Laguna, Tenerife, Spain}
 \affiliation{Departamento de Astrof\'isica, Universidad de La Laguna, E-38206, La Laguna, Tenerife, Spain}
 \email[noshow]{mbianchin.astro@gmail.com}  

\author[0009-0008-2184-1403]{Maitê S. Z. de Mellos}
\affiliation{Departamento de F\'isica, CCNE, Universidade Federal de Santa Maria, Av. Roraima 1000, 97105-900,  Santa Maria, RS, Brazil}
\email[noshow]{maite.mellos@acad.ufsm.br}

 \author[0000-0002-1321-1320]{Rog\'erio Riffel}
 \affiliation{Departamento de Astronomia, IF, Universidade Federal do Rio Grande do Sul, CP 15051, 91501-970, Porto Alegre, RS, Brazil}
 \email[noshow]{riffel@ufrgs.br}  

 \author[0009-0005-0583-5773]{Gabriel L. Souza-Oliveira}
 \affiliation{Departamento de F\'isica, CCNE, Universidade Federal de Santa Maria, Av. Roraima 1000, 97105-900,  Santa Maria, RS, Brazil}
 \email[noshow]{gabrieloliveira2428@gmail.com} 

 \author[0000-0003-1772-0023]{Thaisa Storchi-Bergmann}
 \affiliation{Departamento de Astronomia, IF, Universidade Federal do Rio Grande do Sul, CP 15051, 91501-970, Porto Alegre, RS, Brazil}
 \email[noshow]{thaisa@ufrgs.br}  

\begin{abstract}

We use the \textit{James Webb Space Telescope} (JWST) Mid-Infrared Instrument (MIRI) Medium Resolution Spectrograph (MRS) to investigate the warm molecular and low ionization gas outflows driven by the low-luminosity Active Galactic Nucleus (AGN) in NGC\,3884. By decomposing the emission-line profiles into multiple Gaussian components, we detect both ionized and warm molecular outflows, traced by [Ar\,{\sc ii}]\,6.99\,$\mu$m and H$_2$\,0--0\,S(5), respectively. {The ionized and warm molecular components both extend to similar distances from the nucleus, reaching approximately $\sim$1.2\,kpc.} {In addition to an outflow component,} we detect a double narrow emission component in both ionized and molecular lines, clearly observed along the minor axis with similar amplitudes, while their intensity ratio varies systematically along the major axis. Both components exhibit similar line widths and centroid velocity distributions, indicating that they are not associated with outflows, but instead arise from gas in the galaxy disk. We estimate that the outflow injects momentum into the interstellar medium (ISM) at a rate corresponding to {$\sim$0.015\%} of the AGN bolometric luminosity, consistent with a radiation-driven origin. Assuming a constant outflow rate, the gas reservoir would be depleted on a timescale of {$\sim$26\,Myr}, potentially suppressing nuclear star formation in the host galaxy.
\end{abstract}


\section{Introduction} 

Galactic winds play an essential role in galaxy evolution by regulating the depletion of gas in its host galaxy -- either by transferring momentum or energy to the gas. The production of winds is generally related to high-energy processes such as violent star formation {(SF)}, supernova events and accretion onto super massive black holes (SMBH) -- each with its respective region of influence \citep[e.g.][]{2012ARA&A..50..455F,2017NatAs...1E.165H,2019A&A...632A..26G,2025ApJ...982...28L}. These feedback processes result in multiphase gas, where each phase has distinct properties.

Molecular gas is the main fuel for {SF} and therefore galaxy growth \citep{2020ARA&A..58..157T}, as well as fast growing black holes. Therefore, the analysis of this gas phase gives clear insights to feedback mechanisms acting on galaxies, even at the early stages. The molecular gas is much more complex than its ionized counterpart, and its emission is mainly found spread between the near-infrared (NIR)  to radio wavelengths. For instance, hot molecular gas (T\,$\gtrsim$ 1000\,K) emission {is a significant contributor to} the NIR band {line emission}, as traced by H$_2$ ro-vibrational lines \citep{2013MNRAS.430.2002R,2005MNRAS.364.1041R}. {This is the least abundant molecular gas phase in the ISM}, once H$_2$ is strongly dissociated at temperatures above 2380\,K \citep{1964JChPh..41.3820L}. The Mid/Far-infrared H$_2$ rotational transitions correspond to warm molecular gas with temperatures of a few hundreds of Kelvins; since H$_2$ does not have a permanent dipole moment, its emission becomes very weak below $\sim$100\,K \citep{2013ARA&A..51..207B}. The warm molecular gas (100\,K $\lesssim$\,T\,$\lesssim$ 1000\,K), now accessible through the JWST-MIRI, comprises a valuable dataset to probe the feedback mechanisms acting on galaxies \citep{2022ApJ...940L...5U,2024ApJ...974..127C,2024ApJ...977...36B,2025A&A...693A.321H}.

Warm molecular emission is known to correlate strongly with shocked gas \citep[e.g.][]{Guillard09,Flower10,2010ApJ...724.1193O,Hill14,2023ApJ...951..104A}. Consistently, AGN hosts present higher H$_2$/PAH emission ratios (traced by e.g. the H$_2$ 0-0\,S(3)/PAH\:11.3\,$\mu$m ratio) than those observed in star-forming galaxies \citep{Roussel07,2019MNRAS.487.1823L,Garcia-Bernete24,2025ApJ...982...69R}. In radio galaxies, shocks associated with the radio jet play an important role in the excitation of the gas. In these systems, the H$_2$/PAH ratio is found to be up to two orders of magnitude higher than in normal star-forming galaxies \citep{2010ApJ...724.1193O,2012ApJ...747...95G}. In shock-dominated regions, the population of the H$_2$ energy levels also differs from that produced by other excitation mechanisms, such as UV or X-ray radiation \citep{2008ApJ...678..974N}. In particular, the temperature derived from the upper levels reflects the radiative cooling through higher transitions, as discussed by \cite{2019MNRAS.487.1823L}. Recent JWST observations of nearby AGN hosts reinforce the role of shock excitation in the observed H$_2$ emission, with such shocks being produced by outflows or by the interaction of radio jets with the {ISM} \citep{2024ApJ...977..156D,2025A&A...704A..88L,Delaney25,2025MNRAS.544.3361V,2026A&A...705A..59R,Zhang26}.

The H$_2$ emission is also seen to correlate with optical tracers of shocks in neutral gas, such as [O\:{\sc i}] emission lines, as indicated by single-aperture observations of nearby galaxies \citep{2020MNRAS.491.1518R}. This provides evidence that the highest H$_2$/PAH ratios observed in AGN hosts can be associated with {shocks produced by} massive molecular outflows. 

In this work we use JWST Mid-Infrared Instrument (MIRI) Medium Resolution Spectrograph (MRS) {\citep{2015PASP..127..646W,Labiano21,2023A&A...675A.111A}} observations to map the {kinematics} of the warm molecular and ionized gas in the inner region of the LINER host NGC\,3884, which is the galaxy with the highest H$_2$/PAH ratio among the three selected galaxies from the Blowing star formation Away in AGN Hosts (BAH) sample \citep{2025ApJ...982...69R}. {The BAH project targets three nearby low-luminosity AGNs ($z<0.1$): 3C\,293, NGC\,3884, and CGCG\,012-070, selected for their high H$_2$/PAH and [O\,{\sc i}]/H$\alpha$ ratios, as well as their high [O\,{\sc i}] velocity dispersions.  These galaxies have previously been studied at optical wavelengths using Gemini Multi-Object Spectrograph (GMOS) observations \citep{2023MNRAS.521.3260R,2024MNRAS.528.1476R,2025MNRAS.542.3373R}, revealing ionized-gas outflows in all of them. Near- and mid-infrared observations have also been explored, including studies of the hot and warm molecular outflows in 3C\,293 \citep{2024ApJ...974..127C,2026ApJ...997...40D} and the excess H$_2$ emission in these systems \citep{2025ApJ...982...69R}. NGC\,3884 is a nearby AGN at a redshift of $z=0.023123$ \citep{2016ApJ...819...63R}. Spatially resolved optical observations have revealed ionized-gas outflows extending over several hundred parsecs \citep{2024MNRAS.528.1476R}. The stellar population within the inner kiloparsec is predominantly composed of intermediate-age and old stars ($t > 0.7$ Gyr), but clear signatures of recent star formation ($t \leq 0.7$ Gyr) are also detected in the nuclear region, pointing to a possible rejuvenation episode \citep{Rogerio_BAH5}.}

This paper is organized as follows. In Section~\ref{sec:data} we describe the observations and the emission-line profile fitting procedure. In Section~\ref{sec:results} we present the emission-line flux distributions and kinematics, which are discussed in Section~\ref{sec:disc}. Finally, our conclusions are summarized in Section~\ref{sec:conc}. 
 Throughout this paper, we adopt a distance to NGC\,3884 of 115\,Mpc based on the Tully--Fisher method \citep{tully16}, for which 1\,arcsec corresponds to $\sim$557\,pc.
\begin{figure*}[t]
    \centering
    \includegraphics[width=0.99\linewidth]{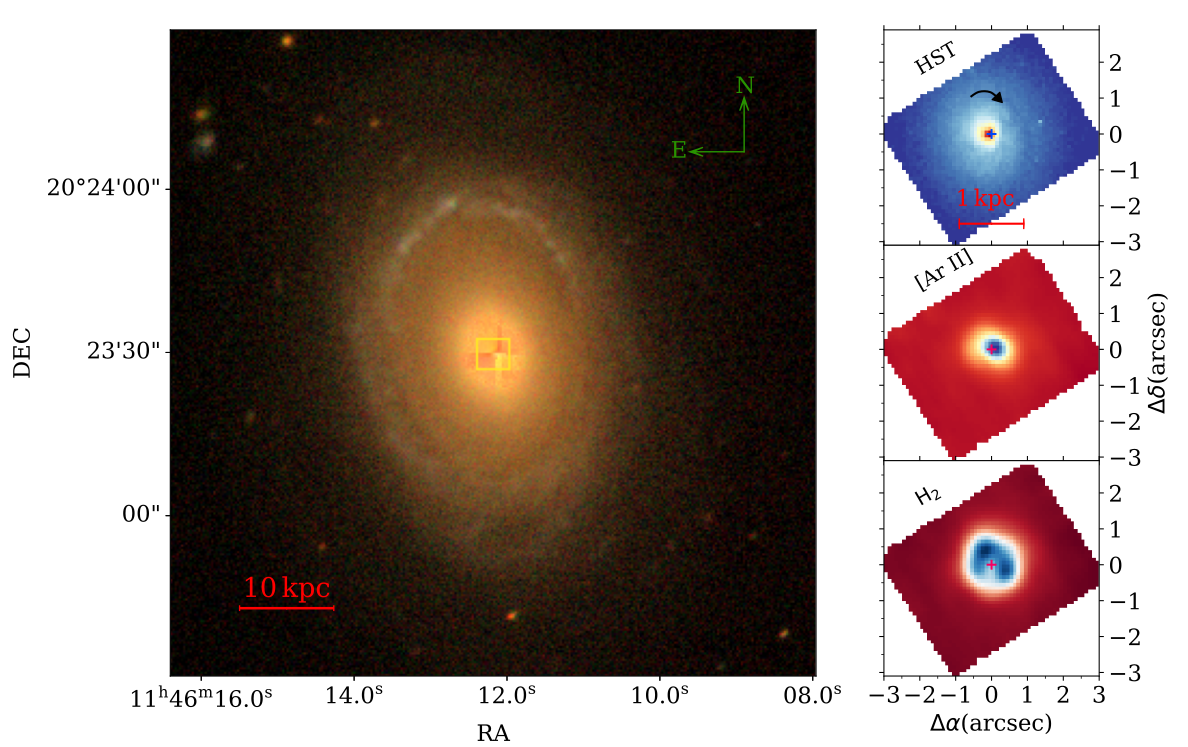}
\caption{Left panel: RGB composition using the Sloan Digital Sky Survey filters $g$, $r$, and $i$; {the central square represents the MIRI–MRS FoV (shown in the right panels). }Right side panels: HST F547M filter centered at $\lambda=5495$\,\AA\ (top), moment 0 (flux) maps for [Ar\,{\sc ii}] (middle), and $\rm H_2\,0$--$0\,S(5)$ (bottom). The cross marks the center of emission. {The arrow at the top panel indicates the position of the structure mentioned in the text.} }
    \label{fig:0}
\end{figure*}

\begin{figure*}[t]
    \centering
    \includegraphics[width=0.99\linewidth,angle=0]{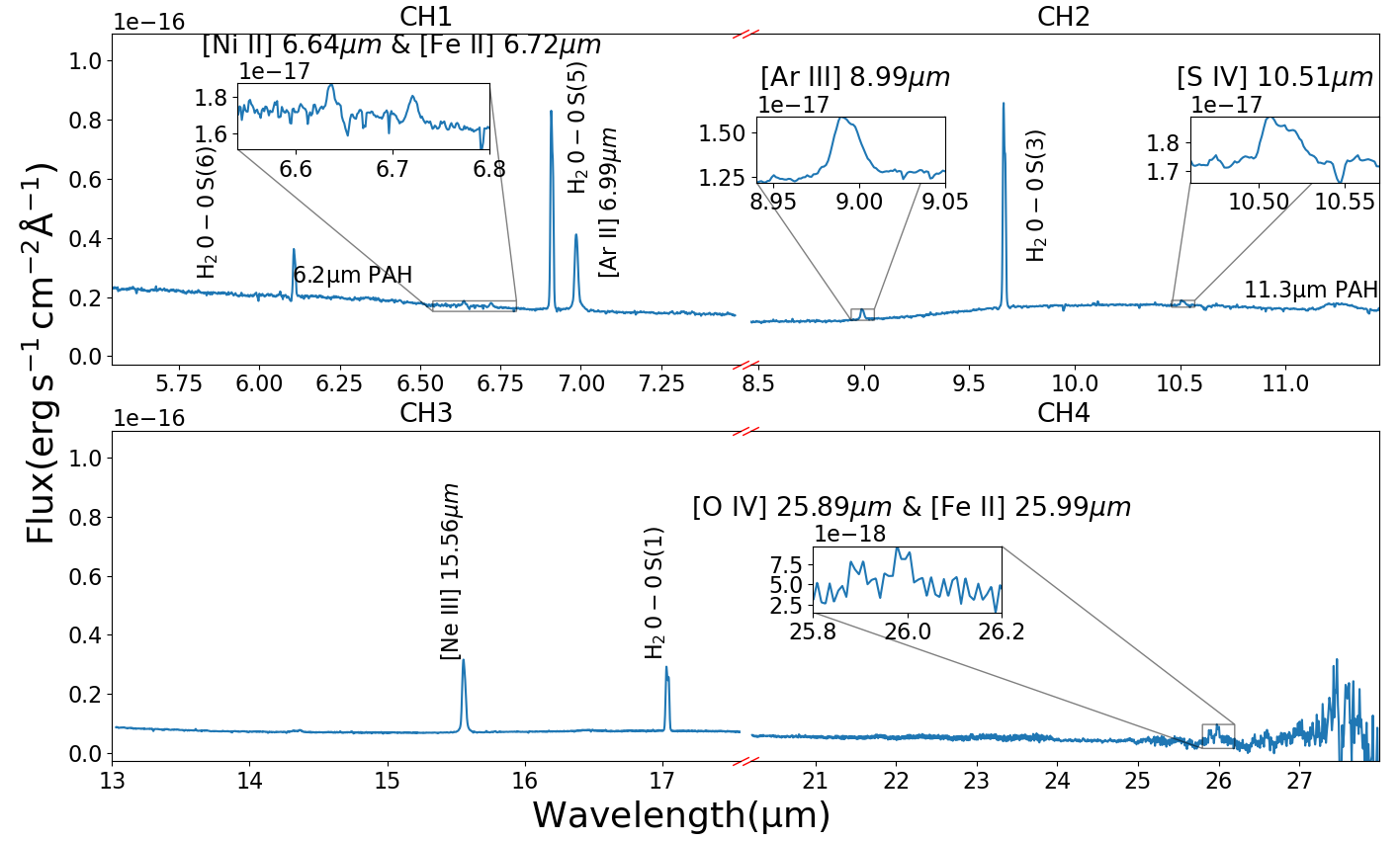}
    \caption{ Spectrum of NGC\,3884 extracted within a radius equal to $2\times$FWHM using a conical aperture (increasing radius with the wavelength{, to account for the growth of the PSF}), centered at the nucleus. Fainter emission lines are highlighted in zoomed-in windows. Each channel is labeled above the plot, while the gaps between sub-bands are omitted.}
    \label{fig:spec}
\end{figure*}

\section{Data and Measurements} \label{sec:data}

We analyze JWST MIRI MRS observations of NGC\,3884 obtained in the medium and long sub-bands as part of a Cycle~1 JWST program (Prop.\ ID: 1928). The observations were carried out using a four-point dither pattern, with 30 groups and a single integration per dither position in the slow readout mode, summing to a total on target exposure time of 57\,min and 20 sec. The data were processed using the JWST Science Calibration Pipeline \citep{2024zndo..12692459B}, version 1.13.4, with the reference file \texttt{jwst\_1188.pmap}. After the standard reduction, we applied a custom cleaning pipeline designed to remove spurious data points that remained after the main pipeline processing. We performed the background subtraction using a 1D master background spectrum, extracted from the background exposures. The resulting data partially cover the spectral range from 5.6\,$\mu$m to 27.8\,$\mu$m, missing wavelength windows corresponding to the short sub-band (7.5--8.5\,$\mu$m, 11.4--13.0\,$\mu$m, and 17.6--20.2\,$\mu$m). {We performed the cube build multi-mode to assembly the individual channel in a larger cube featuring the whole spectral extent (5.6-27.8\,$\mu$m)}. The spatial resolution in our data ranges from $\sim$190\,pc {(0.34\arcsec)} at 7\,$\mu$m to $\sim$370\,pc {(0.67\arcsec)} at 17\,$\mu$m, as derived from the relation presented by \cite{Law23} for the point spread function {(PSF)} full width at half maximum (FWHM). We also use Hubble Space Telescope (HST) photometric data of the filter F547M, which was taken under the proposal ID 6837; available through HST archive. 

Figure \ref{fig:0} shows a large-scale image of NGC\,3884, with the {MRS channel 1} field of view {(FoV)} of our observations overlaid. {We used the \texttt{reproject\_interp} routine from the \texttt{astropy} package \citep{2013A&A...558A..33A} to reproject the HST image shown in the first panel of the right-hand column of Fig.~\ref{fig:0} onto the spatial grid of the MIRI Channel 1 maps, ensuring that it covers the same {FoV} and facilitates a direct spatial comparison.} {MIRI-MRS FoV scales with each channel as follows: CH1 (3.2''$\times$3.7''), CH2 (4.0''$\times$4.8''), CH3 (5.2''$\times$6.2'') and CH4 (6.6''$\times$7.7''). The spectra of NGC\,3884 show strong and extended H$_2$ emission across nearly the entire MIRI MRS FoV (Fig. \ref{fig:spec} and \citealt{2025ApJ...982...69R}).}
The [Ne\,{\sc iii}] 15.56\,$\mu$m emission line shows strong and extended emission, {while} species such as [Ar\,{\sc ii}] 6.99\,$\mu$m, [Ar\,{\sc iii}] 8.99\,$\mu$m and [Fe\,{\sc ii}] 25.99\,$\mu$m are also detected, although their emission is more concentrated toward the nucleus. {Other species} such as [S\,{\sc iv}] 10.51\,$\mu$m and [O\,{\sc iv}] 25.89\,$\mu$m, are detected but are extremely faint, and no coronal emission is observed. {Kinematic and flux maps for all line emission found in BAH sample will be included in a forthcoming paper}. PAH emission in NGC\,3884 extends up to $\sim$2$\arcsec$ partially aligned to the North-South {direction} \citep{2025ApJ...982...69R}.  We show the integrated spectrum of NGC\,3884 in Fig.~\ref{fig:spec}. The spectrum was extracted using a conical aperture centered at the peak of {continuum} emission at {7$\mu$m}, with a radius equal to twice the FWHM of the PSF, given by FWHM$\approx 0.033\lambda + 0.106$ \citep{Law23}. The fainter emission lines are highlighted in zoomed-in panels.  

\subsection{Emission Line Fitting}

{We start by accounting for the continuum emission following the procedure described by \citet{2024ApJ...974..127C}. For each spaxel, we construct the observed continuum from 5.6 to 27.8\,$\mu$m by applying a double LOWESS smoothing to the observed spectrum, using 30\% and 10\% of the data points in succession, followed by linear interpolation.  The emission lines are then fitted in the continuum-subtracted data cube using the \texttt{ifscube} package, version 3.11 \citep{2021MNRAS.507...74R}.} The code allows the use of Gaussian profiles or Gauss--Hermite series to model the emission-line profiles and enables the application of constraints to the fit parameters. It can be applied both to single spectra and to spectral cubes. {We simultaneously fitted all the detected H$_2$ lines with Gaussian profiles, namely H$_2\,0$--$0\,S(1)$, H$_2\,0$--$0\,S(3)$, H$_2\,0$--$0\,S(5)$, and H$_2\,0$--$0\,S(6)$, which trace the warm molecular gas, as well as the [Ar\,{\sc ii}] 6.99\,$\mu$m line tracing the low-ionization gas, following a procedure similar to that described by \citet{2024ApJ...974..127C}. {The emission lines of the two gas phases were fitted independently, with no kinematic parameters tied together.}}

To properly reproduce the observed emission-line profiles in NGC\,3884 {(see Fig. \ref{fig:gaus})}, three Gaussian components are required for each emission line: two components to reproduce the disk emission, as the rotation axis presents a clear double-peaked profile, and an additional broader component to account for the turbulent gas associated with outflows. We set the \texttt{refit} parameter of IFSCUBE to use the best-fitting parameters from spaxels located within 3 pixels of the target spaxel as initial guesses for the next fit. This allows the routine to preserve the spatial continuity of the fitted parameters. For example, if one of the Gaussian components has zero amplitude in the neighbouring spaxels, the code uses the parameters of the remaining two Gaussian components as initial guesses for the next fit, and similarly for other cases. We adopted three Gaussian components as the maximum number of components in the final fit. {Only Gaussian components with amplitudes greater than $5\sigma$ of the local noise are considered significant.} As a sanity check, we verified that the resulting number of significant components is consistent with the Akaike Information Criterion \citep[AIC;][]{Akaike1974} comparison between models with one, two, and three Gaussian components.  We assign each of the narrow Gaussian components to one side of the disk, allowing their parameter space to overlap around zero velocity. The broader Gaussian component, on the other hand, is allowed to vary freely, although its velocity dispersion is constrained to be larger than that of the narrow components ({Fig.~\ref{fig:gaus}}). 

In Fig.~\ref{fig:gaus} we present the H$_2\,0$--$0\,S(5)$ velocity field along with examples of the emission-line fits. The velocity field is derived using {the flux weighted sum of velocity components}, computed by weighting each component by its flux.  The H$_2$ velocity field shows a clear rotation pattern, with the southern side of the disk approaching and the northern side receding, similar to that observed for the stars and ionized gas \citep{2024MNRAS.528.1476R}. {In Fig. \ref{fig:gaus}} we show examples of fits of the H$_2$0-0\,$S$(5) emission-line profile for five positions: the nucleus, two positions along the major axis and two positions along the minor axis.  
{The two narrow components are detected predominantly along the galaxy's minor axis, with comparable amplitudes. The spatial distribution of their relative flux ratio is presented in Appendix~\ref{sec:appendixA}. Along the major axis, however, the ratio between the blue and red components varies systematically. } 
 
{In addition, the blue- and redshifted narrow components show similar mean widths and amplitudes (see Appendix~\ref{sec:appendixB}). } A similar behavior is also observed {for} the ionized gas emission lines {in NGC\,3884 MIRI spectra} {(Costa-Souza et al., in prep.)} and is consistent with results found for a sample of 35 double-peaked narrow emission-line galaxies observed as part of the Mapping Nearby Galaxies at Apache Point Observatory (MaNGA) survey, where the double-peaked profiles are associated with emission from the galaxy disks \citep{2025MNRAS.543.3089Z}. {{In the latter}, the disk rotation is producing a Doppler shift in the disk emission-line component {unresolved around the zero velocity line}. Thus, we attribute the two narrow components observed in NGC\,3884 to the galaxy disk. In the following section, we present the flux and kinematic distributions combining these two components to represent the overall narrow-line emission. {The broad component is primarily observed in the central region and is attributed to the warm molecular phase of the outflow, as previously reported for the ionized gas by \citet{2024MNRAS.528.1476R}.

\begin{figure*}[]
    \centering
    \includegraphics[width=0.99\linewidth]{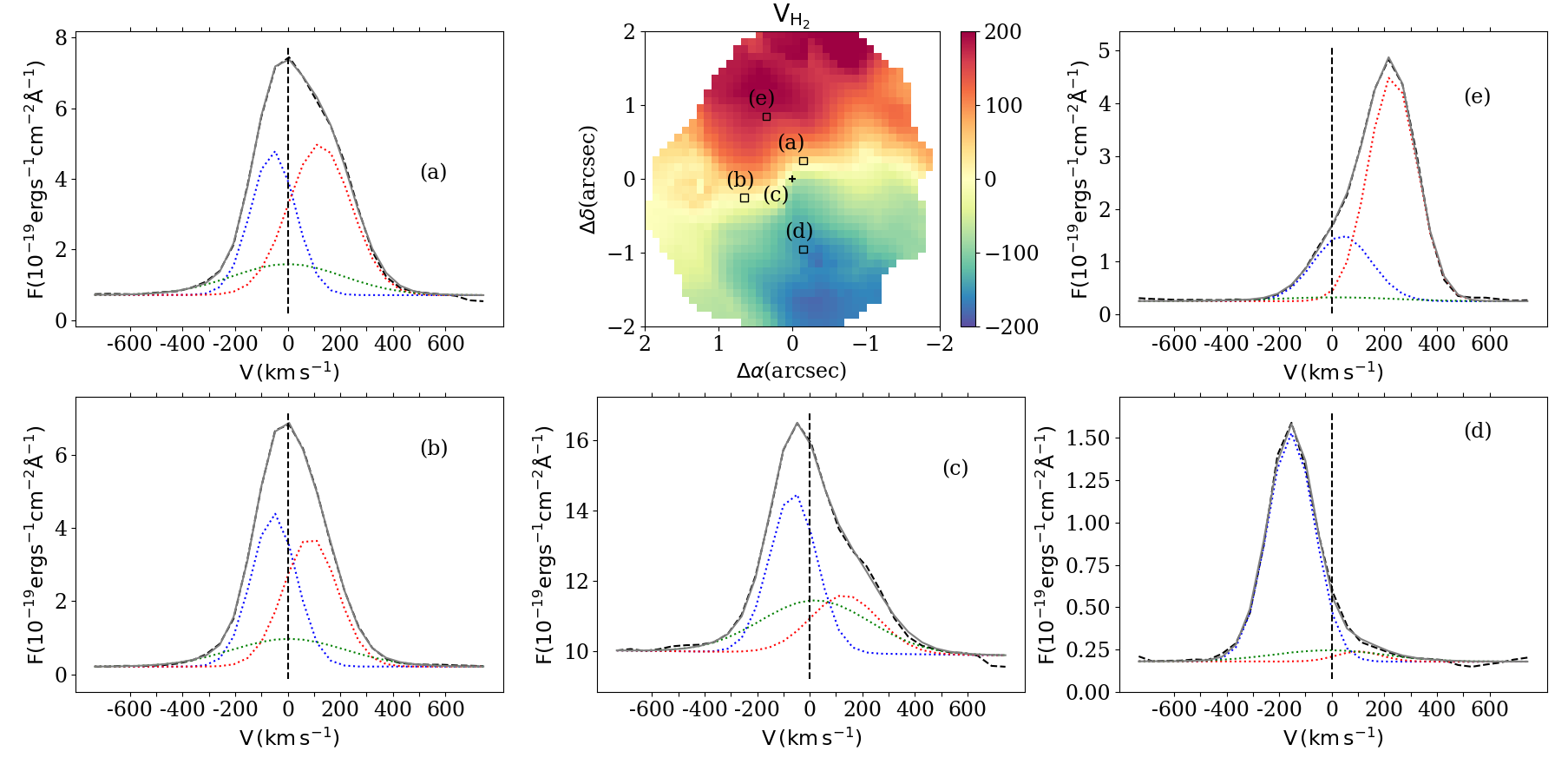}
    \caption{Example of H$_2$\,0--0\,$S$(5) line spectra at the emission center and along the major and minor axes. The observed profiles are shown as dashed black lines, while the best-fitting model is displayed as a solid gray line. The narrow components are represented by blue and red dotted lines, and the broad component by a green dotted line. {The top-center panel shows the centroid velocity obtained by combining the two narrow Gaussian components, representing the disk}. The black squares indicate the positions of the spectra extracted along the minor axis and major axis. The {nucleus} is indicated by a black cross.}
    \label{fig:gaus}
\end{figure*}

\section{Results} \label{sec:results}

We use the MIRI MRS Channel~1 long sub-band to explore the emission of NGC\,3884{. This} channel has the highest spatial resolution among the MIRI–MRS channels (see Sect. \ref{sec:data}), which motivated the choice of the emission line set we are working on. At the galaxy redshift ({z = 0.023123}), this band covers the H$_2\,0$--$0\,S(5)$ emission line as well as the [Ar\,{\sc ii}] 6.99\,$\mu$m (hereafter [Ar\,{\sc ii}]) ionized-gas emission line. {We show the [Ar {\sc ii}] and H$_2$ flux distributions at the insets of Fig.~\ref{fig:0}, obtained by taking the moment-0 of the line emission}. The spatial distributions of the ionized and molecular gas are quite distinct. The molecular gas presents a ring-like morphology, apparently tracing a nuclear ring -- \cite{Rogerio_BAH5} report an excess of stars with ages ($\tau$) within 0.70 Gyr $<$ $\tau$ $\leq$ 2.0 Gyr, matching the ring position. This structure is not observed in other emission lines, but only in the molecular H$_2$ transitions. {Furthermore, the ring structure present two regions of intense emission (1) at north-east of the nucleus, (2) at south-west; and a region at south, with a fainter emission}. In contrast, the low-ionization [Ar\,{\sc ii}] emission peaks at the center and becomes progressively fainter toward the outer regions. 
{In Fig.~\ref{fig:0}, we also show a HST image obtained with the F547M filter, where a slightly fainter structure is visible toward the western side of the nucleus{, as indicated by the black arrow}. This feature may be associated with a dust lane, although its morphology is not sufficiently well defined to support a firm identification. It appears as a localized decrease in the continuum flux and is not evident in the emission-line flux distributions.}

As mentioned above, the two narrow components are associated with emission from the galaxy disk. Therefore, we combine them to produce single flux and kinematic maps. The flux distributions are obtained by summing the fluxes of the two narrow components, while the velocity field is defined as the centroid velocity {($v_{c}$)} (i.e., {the flux weighted velocity}). The velocity width is measured using the $W_{80}$ parameter, defined as the width of the line encompassing 80\% of its total flux, considering both narrow components. The results for the outflow component are derived from the broad Gaussian fit. In Fig.~\ref{fig:maps}, we present the flux, centroid velocity, and $W_{80}$ maps for the H$\rm_2\,S(5)$ and [Ar\,{\sc ii}] emission lines, showing both the disk and outflow components. The velocity fields are shown after subtracting the systemic velocity of the galaxy, defined as the mean {$v_{c}$} of the H$_2$ disk component within the inner 1\,arcsec radius region. Gray regions in these maps indicate locations where the amplitude-to-noise ratio ($A/N$) is less than 5, with the noise level estimated as the standard deviation of the continuum adjacent to each emission line. Maps for the H$_2\,S(1)$, S(3), and S(6) emission lines are presented in Appendix~\ref{sec:appendix}, showing flux distributions and kinematics similar to those of the H$_2\,S(5)$ line.

In addition to the prominent H$_2$ ring, we detect extended H$_2$ emission out to $\sim$2\,arcsec from the nucleus, covering nearly the entire MIRI MRS field of view. The H$_2$ velocity field for the disk component displays a clear rotation pattern, with a projected amplitude of $\sim$150\:km\:s$^{-1}$ and kinematic major axis oriented approximately along the north-south direction. Some deviations from a purely rotating disk are observed, such as distortions in the zero-velocity line. The $W_{80}$ values for the H$_2$ disk component are below $\sim$400\,km\,s$^{-1}$ in most regions, with the lowest values ($\sim$200\,km\,s$^{-1}$) observed along the major axis, to the south and north of the nucleus.

The [Ar\,{\sc ii}] flux distribution for the disk component peaks at the galaxy nucleus and decreases with distance, lacking the ring-like morphology seen in the H$_2$ emission. Its velocity field exhibits a similar rotation pattern to that of H$_2$, but with more distorted isovelocity lines, indicating a more perturbed gas. This is further supported by {higher $W_{80}$ values for [Ar\,{\sc ii}] ($\langle\rm W_{80} \rangle$ = 311$\rm\,km\,s^{-1}$) compared to those of H$_2$ ($\langle\rm W_{80} \rangle$ = 240$\rm\,km\,s^{-1}$).} In addition these isovelocity distortions indicate that, NGC\,3884 may have underlying hidden structures or secondary outflow components.

The outflow component is more restricted to the central region, with flux distributions more elongated along the east-west direction up to $\sim$1.5 arcsec from the nucleus for both {H$_2\,0$--$0\,S(5)$} and [Ar\,{\sc ii}]. {The outflow component peaks slightly southwest of the nucleus, at a projected distance of $\sim0.1$--$0.2''$. \citet{2024MNRAS.528.1476R} found the outflow to peak at the nucleus; however, their observations have substantially lower spatial resolution, making such a small displacement difficult to resolve. Moreover, the displacement observed here is smaller than the PSF of our observations and should therefore be interpreted with caution.} The velocity map for the broad component shows slightly blueshifted velocities ($\sim$50\,km\,s$^{-1}$) over most regions, with some locations exhibiting higher blueshifts of up to $\sim$150\,km\,s$^{-1}$ on the eastern side. The velocity dispersion for the H$_2$ broad component ranges from $400 \lesssim W_{80} \lesssim 600$\,km\,s$^{-1}$, while larger values are observed for [Ar\,{\sc ii}], ranging from $700 \lesssim W_{80} \lesssim 900$\,km\,s$^{-1}$, indicating that the ionized gas is tracing a more turbulent phase of the outflow.  {The [Ar\,{\sc ii}] outflow velocities fall within the same range as those reported by \cite{2024MNRAS.528.1476R} using optical ionized-gas tracers, although their flux distributions differ, with the optical emission showing a more rounded morphology. Such differences are not unexpected, as these lines trace gas under different excitation conditions and may also be affected differently by the relative abundances of the emitting species, and therefore do not necessarily originate from the same spatial regions. In addition, the observations of \cite{2024MNRAS.528.1476R} are seeing-limited, with a spatial resolution of $\sim0.85''$, whereas our JWST/MIRI observations provide a substantially higher resolution. }

In summary, our results show that both the warm molecular and ionized gas within the inner 1.7\,kpc radius of NGC\,3884 exhibit three kinematic components: two associated with gas in the plane of the galaxy disk, and another associated with outflows driven by its low-luminosity AGN. In the next section, we focus on the warm molecular gas component, determining its temperature distribution and estimating its outflow properties, following the discussion presented in \citet{2024ApJ...974..127C} for the radio-loud AGN host 3C\:293.

\begin{figure*}[t]
    \centering
    \includegraphics[width=0.9\linewidth]{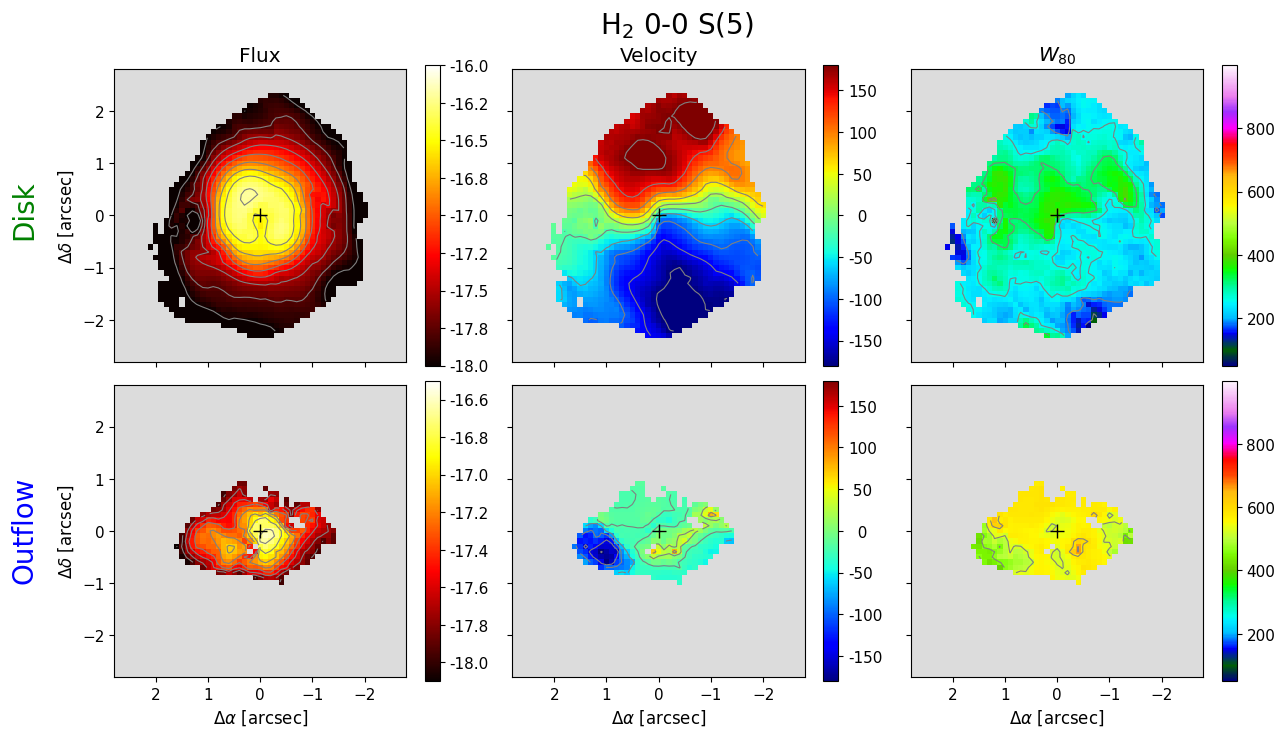}
    \includegraphics[width=0.9\linewidth]{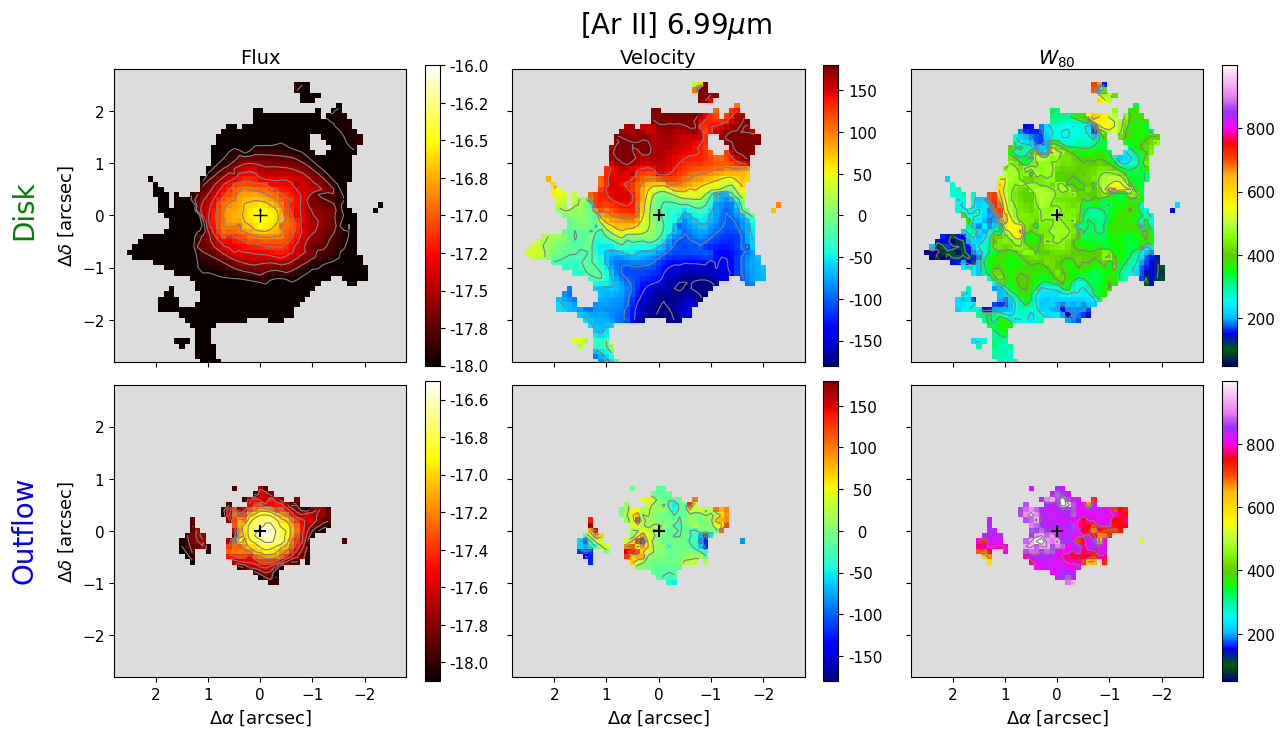}
    \caption{Flux and kinematic maps for the H$_2\,0$--$0\,S(5)$ {emission}, {upper block;} and [Ar\,{\sc ii}] emission, {lower block}. The flux distributions are shown in the left panels, with the color bar indicating flux values in logarithmic units of erg\,s$^{-1}$\,cm$^{-2}$. The centroid velocities are shown in the middle panels in km\,s$^{-1}$, after subtraction of the systemic velocity of the galaxy. The right panels display the $W_{80}$ maps in km\,s$^{-1}$. Gray regions correspond to masked locations where the emission lines are not detected with $A/N > 5$, and the central crosses mark the position of the galaxy nucleus, defined as the peak of the continuum emission. {The contours are designed to highlight the details in each of the maps.} The disk and outflow components are labeled in each row.}
    \label{fig:maps}
\end{figure*}

\section{Discussion} \label{sec:disc}

\begin{figure}[t]
    \centering
    \includegraphics[width=0.99\linewidth]{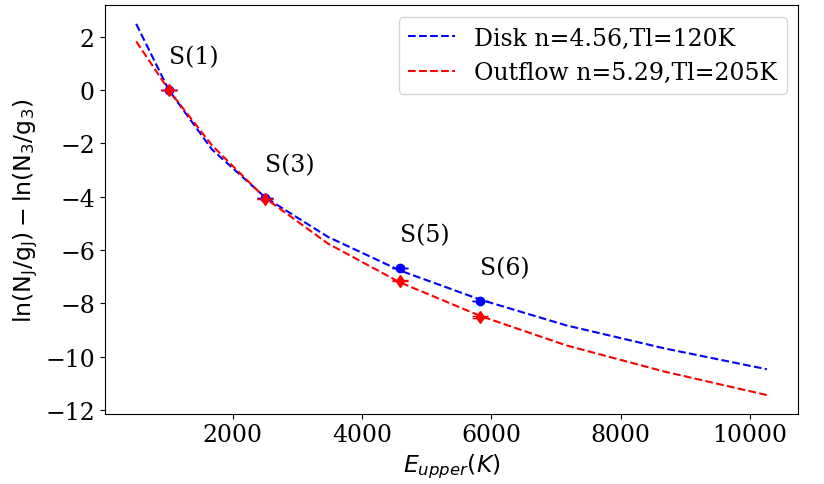}
    \caption{Excitation diagram for the disk  and outflow components. The markers show the data points while the lines display the fitted model; blue for the disk component and red for the outflow component. The plot shows the natural logarithm of the column density of each level divided by the degeneracy of the state, normalized by the level J=3, as a function of the energy of the upper level in {Kelvins}.}
    \label{fig:pl}
\end{figure}

\subsection{H$_2$ excitation diagrams and gas masses}
\label{sec4.1}

Measuring the temperature of H$_2$ gas is essential for determining other {quantities}, such as the molecular gas mass and {mass} outflow rate. Molecular gas is found with a wide range of temperatures in galaxies with varying rates of star formation and varying levels of AGN activity \citep{Roussel07,2006ApJ...648..323H,2010ApJ...724.1193O}. 

One approach to measure the {molecular gas} properties is to describe the gas using a distribution of excitation temperatures. {
The temperature distribution of warm molecular H$_2$ is often approximated by a power law,
$dN \propto T^{-n}\,dT$.
This parametrization can describe molecular gas spanning a range of temperatures, such as gas heated by shocks and subsequently undergoing radiative cooling \citep{1979ApJS...41..555H,1989ESASP.290..281B,2008ApJ...678..974N}. }
This method is particularly powerful, as it can provide insights into the mechanisms responsible for molecular gas excitation \citep{2017ApJ...836...76A}. The full modeling procedure is described in \citet{2024ApJ...974..127C}. In short, the column density of a given rotational state $J$ can be expressed as
\begin{equation} \label{eq:NJ}
    N_{J} = \int_{T_l}^{T_u} f(T)\, m\, T^{-n}\, dT,
    \label{eq:1}
\end{equation}
where $f(T)$ accounts for the Boltzmann factor of the ortho states ({i.e.,} odd $J$), and $m$ is a normalization constant that depends on the upper ($T_u$) and lower ($T_l$) temperatures of the model, the power-law index ($n$), and the total H$_2$ column density ($N_T$), given by
\begin{equation}
    m = \frac{N_T (n-1)}{T_l^{1-n} - T_u^{1-n}}.
\end{equation}
$N_J$ can also be calculated as
\begin{equation}
    N_J = \frac{4 \pi F_J \lambda}{\Omega\, h c\, A_{u-l}},
\end{equation}
where $F_J$ is the flux of the line at the $J$ state and wavelength $\lambda$, $\Omega$ is the solid angle, $A_{u-l}$ is the Einstein coefficient of the transition from $J+2$ to $J$, $h$ is Planck's constant, and $c$ is the speed of light.  

By combining this expression with Eq.~\ref{eq:NJ}, we obtain
\begin{equation}\label{eq:NT}
    N_T = \frac{4 \pi F_J \lambda \left(T_l^{1-n} - T_u^{1-n}\right)}{\Omega\, h c\, A_{u-l} (n-1) \int_{T_l}^{T_u} f(T)\, T^{-n}\, dT}.
\end{equation}

To reduce the number of free parameters, an upper temperature, $T_u$, can be fixed to a value appropriate for the data. Following \cite{2024ApJ...974..127C} and \cite{2011ApJ...726...76Y}, we adopt $T_u = 5000$\,K. {This upper limit encompasses both the warm and hot molecular gas regimes, which we therefore distinguish when interpreting the resulting temperature distribution. } To determine $n$ and $T_l$, we fit the column densities of each transition {(measured by spatially integrating the emission-line flux separately for the disk and outflow components)} as a function of the upper-state temperature {using the following relation, obtained normalizing Eq. \ref{eq:1} with the H$_{2}$\,0-0\,S(1) transition (i.e. J=3)}

\begin{equation}
    \frac{N_{J}/g_{J}}{N_{3}/g_{3}} = \frac{\int_{T_l}^{T_u} e^{-E_J / k_B T}\, T^{-n}\, dT}{\int_{T_l}^{T_u} e^{-E_3 / k_B T}\, T^{-n}\, dT},
\end{equation}
where $E_J$ is the energy of the $J$ level and $g_J$ is its statistical weight.

{The H$_2$ excitation diagram provides a useful diagnostic of the excitation conditions and temperature distribution of the observed molecular gas. It relates the normalized column density of molecules in each upper energy level (y-axis) to the excitation temperature of that level (x-axis). The resulting distribution can be used to identify the temperature regimes populated by the warm molecular gas and to characterize its excitation conditions.} In Fig.~\ref{fig:pl}, we show the H$_2$ excitation diagram for both the disk and outflow components, along with their corresponding best-fit power-law temperature distributions, {yielding lower temperatures of 120$\pm\,60$\,K (205$\pm\,15$\,K) and power-law indexes of 4.56$\pm\,0.13$ (5.29$\pm\,0.11$) for the disk (outflow)} component. The power-law index ($n$) can be used to investigate the H$_2$ excitation. Shock-heated molecular gas is expected to show lower indexes, indicating higher temperatures and a larger fraction of hot gas compared to photodissociation regions \citep{togi16}. The power-law indexes {reported in the literature} show significant variation among different studies, with values ranging from 3.79--6.39 for galaxies in the Spitzer Infrared Nearby Galaxies Survey {\citep{togi16}, $\sim$4.2--5.0 in Stephan’s Quintet \citep{2017ApJ...836...76A,Appleton23}, and 2.5--5.0 in ultra-luminous infrared galaxies \citep{zakamska10}, with the lowest values observed in shock regions}. 

The $n$ values derived for both the disk and outflow components of NGC\,3884 fall within the range reported in the literature. However, we find a {steeper temperature distribution for the outflow component compared to the disk, as indicated by its relatively higher $n$ value. In the adopted power-law temperature distribution, larger $n$ values correspond to steeper distributions and a relatively larger contribution from lower-temperature gas, whereas smaller $n$ values indicate a flatter distribution with a relatively larger contribution from higher-temperature gas. Thus the outflow component displays lower overall} temperatures when compared to the disk.  \citet{2025ApJ...982...69R} applied {a similar} fitting procedure, but separating a kinematically disturbed region from a virial-dominated region using the H$_2$ S(3) line width as a reference, {instead of separating outflow and disk components}. They found similar slopes for the power-law relation in both regions.  This {different results, resulting from the different approaches} suggests that, although molecular outflows are present, shocks associated with them are likely not the unique H$_2$ excitation mechanism at play. On the other hand, the $T_l$ value for the outflow component is higher than that of the disk, indicating that the outflowing gas is hotter than the gas in the galaxy disk. Therefore, heating by lower velocity shocks related to the outflow cannot be ruled out, and the most likely scenario for the H$_2$ excitation in NGC\:3884 is a combination of different processes, including shocks driven by AGN outflows and heating by AGN radiation, as reported for other nearby AGN hosts \citep[e.g.][]{Delaney25,Delaney26,Zhang26}. An important caveat is that our data include only four H$_2$ pure rotational transitions and, in particular, the H$_2$ 0--0 S(0) transition, which provides the main constraint on the coldest molecular gas component, {falls outside the spectral coverage at NGC\,3884 redshift.}

Finally, we can calculate the molecular gas mass by
\begin{equation} \label{eq:m}
    M=(4/3) m_{\rm H_{2}} N_{T} \Omega D^{2},
\end{equation}
where $m_{\rm H_{2}}$ is the H$_2$ mass and $D$ is the distance to the galaxy. 
We adopt an ortho-to-para ratio of OPR = 3, as we are unable to derive a reliable estimate from the limited number of detected transitions. The (4/3) scale factor accounts for the para-H2 contribution. We use the H$_2$\,0--0\,S(1) emission line as a reference to estimate the masses and outflow properties, owing to its high intensity and larger spatial extent. {It also probes lower excitation temperatures, tracing a larger fraction of the warm molecular gas, and lies in a region of high MIRI/MRS quantum efficiency. In contrast, S(5) and S(6) have a smaller spatial coverage, while S(3) may be more affected by uncertainties in the extinction correction due to its proximity to the silicate absorption feature.}

Using the $T_l$, {$T_u$\,=\,5000\,K} and $n$ values to derive $N_T$ from Eq.~\ref{eq:NT}, and substituting it into Eq.~\ref{eq:m}, we obtain a warm molecular hydrogen mass in the disk of $\rm M_{disk} = (4.12\pm0.05)\times10^{7}\,M_{\odot}$, and a mass of $\rm M_{outflow} = (2.70\pm0.12)\times10^{6}\,M_{\odot}$ for the outflow component. Thus, the outflow carries about 6\% of the total mass of warm molecular gas in the inner region of NGC\:3884. For comparison, typical values of hot molecular ($T\sim2000$~K; traced by near-IR H$_2$ lines) gas masses in the inner $\sim$kpc of nearby AGN hosts are in the range $10^2$--$10^4$~M$_\odot$ \citep[e.g.][]{2021MNRAS.504.3265R}, while cold molecular gas ($T\sim100$~K; traced by CO transitions) accounts for $10^5$--$10^8$~M$_\odot$ \citep[e.g.][]{2013MNRAS.428.2389M}. Thus, the derived warm molecular gas mass for NGC~3884 lies between the typical masses observed for hot and cold molecular gas, which may suggest that the molecular gas spans a continuous range of temperatures between these two regimes.

 \subsection{Outflow properties}

Here we estimate the warm molecular gas outflow properties using the measurements of the broad emission-line component. The elongated emission-line flux distribution observed for this component {(extending to up to $\sim$1.2\,kpc, see Fig. \ref{fig:maps})} may indicate a bipolar outflow partially aligned with the line of sight, with the eastern side pointing towards us, consistent with the blue-shifted gas observed on that side. On the other hand, the line width decreases toward the edges of the outflow, while the velocities remain mostly slightly blue-shifted, consistent with an expanding shell. We adopt a spherical geometry to estimate the outflow properties driven by the AGN. However, given the difficulty in determining the exact geometry of the outflows in NGC\,3884, an additional uncertainty arises from the choice of geometry, in particular the differences between spherical and conical models (see, e.g., the different approaches to measuring outflow properties in \citealt{2020A&A...633A.134L}).

Henceforth, assuming that the outflows have been expelled at a constant mass outflow rate{, t}he mass flux through a spherical shell of thickness $\Delta R$ can be estimated using the following equation:
 \begin{equation}
    \dot{M} = \frac{M_{out,\Delta R}V_{out,\Delta R}}{\Delta R},
\end{equation}
where $M_{out,\Delta R}$ is the mass of gas within the boundaries of each shell; and $V_{out,\Delta R}$ is the propagation velocity of the shell. We define the velocity of the outflow for each shell as {$V_{out,\Delta R} = |\langle V_{\rm wind}\rangle_{\Delta R}| + 2\langle \sigma_{\rm wind}\rangle_{\Delta R}$,  similar to definitions widely used as a proxy for the outflow velocity in previous studies \citep{2013ApJ...768...75R,2013MNRAS.436.2576L,2019MNRAS.488.4126P,2021MNRAS.504.3890D,2024ApJ...974..127C}}. This definition provides a more reliable tracer of the outflow velocity than the centroid velocity, $W_{\rm 80}$, or a combination of both, as it accounts for the high-velocity gas at the edges of the shells. 

To compute the outflow properties, we define concentric shells of thickness $\Delta R = 0.3\arcsec$, centered on the galaxy nucleus. This choice is comparable to the MIRI/MRS PSF at the H$_2$ S(1) wavelength, ensuring that the innermost aperture encompasses a substantial fraction of the unresolved nuclear emission. {Since we haven't corrected for the PSF emission, the subsequent shells still carries a very small flux contamination stemming from the point source emission.} {Since the molecular gas at temperatures above 1000\,K is more directly comparable to the gas traced by near-infrared H$_2$ ro-vibrational transitions, we separate the molecular gas into two temperature regimes from this point onward: warm ($T_{\rm l}\leq T < 1000$\,K) and hot ($1000\leq T\leq5000$\,K). This distinction allows us to compare the mass outflow rates of the hot component with those derived from near-infrared H$_2$ tracers.} The gas masses are then calculated for the temperature ranges: the warm molecular gas (204--1000~K) and the hot molecular gas (1000--5000~K) {over the entire FoV, then we use a masks to only get the mass and velocity inside each of the shells \citep[for further details see][]{2024ApJ...974..127C}}. The top panel of Fig.~\ref{fig:mout} shows the radial profiles of the mass-outflow rate for the warm (blue triangles) and hot (orange squares) molecular phases. {The uncertainties in the derived gas masses are estimated by propagating the uncertainties in the measured line fluxes. Assuming a Gaussian noise distribution in the continuum, the uncertainty in the flux is approximately estimated as $\sigma_{\rm cont}\times\mathrm{FWHM}$, where $\sigma_{\rm cont}$ is the standard deviation measured in a region close to the emission line and FWHM is the full width at half maximum of the line. These flux uncertainties are then propagated to the corresponding gas masses. The uncertainties in the derived outflow rates are expected to be larger {than the ones in flux measurements}, as these quantities also depend on the outflow velocity and the assumptions adopted regarding its geometry.}  The radial profiles show a {constant} mass outflow rate {near the nucleus}, followed by a nearly linear decline beyond $\sim$500\,pc from the emission center, with the warm molecular gas accounting for {closer to one hundred percent} of the total {molecular} outflow rate (warm+hot components). The outflow rates reach peak values of $\dot{M}_{\rm warm} = 1.7 \pm 0.02\,M_{\odot}\,{\rm yr}^{-1}$ and {$\dot{M}_{\rm hot} = (1.89 \pm 0.002)\times10^{-3}\,M_{\odot}\,{\rm yr}^{-1}$}. {Both profiles show a relatively flat distribution within the inner $\sim1''$, forming a plateau around these peak values.}
 
\begin{figure}[t]
    \centering
    \includegraphics[width=0.99\linewidth]{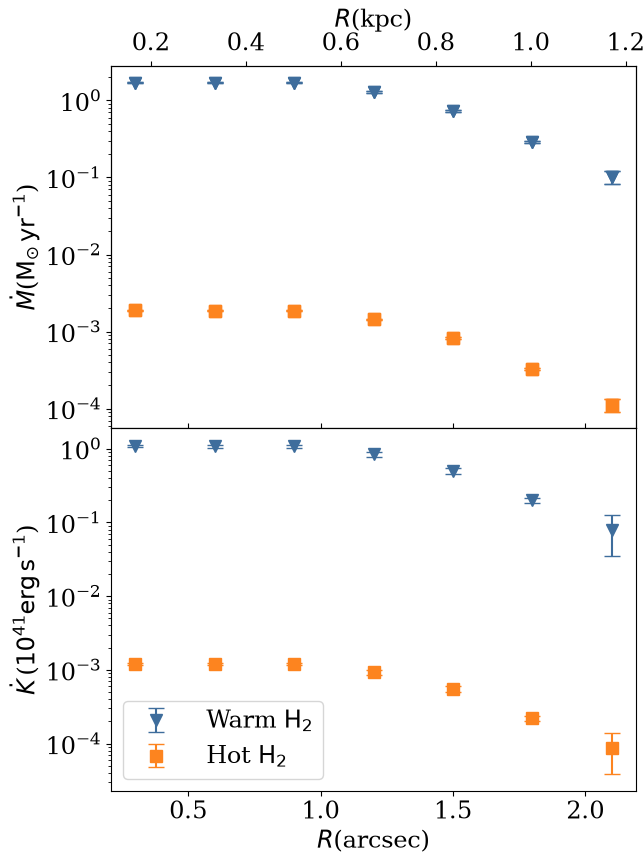}
    \caption{Outflow properties of NGC\,3884: mass outflow rate in the top panel and kinetic power in the bottom panel for the warm (blue triangles) and hot (orange squares) molecular gas components.}
    \label{fig:mout}
\end{figure}

We can also estimate the kinetic power of the outflows, using:

\begin{equation}
    \dot{K} = \frac{M_{out,\Delta R}(V_{out,\Delta R})^{3}}{2\Delta R}.
\end{equation}

 The bottom panel of Fig.~\ref{fig:mout} shows the radial profiles of the kinetic power, $\dot{K}$, for the warm and hot molecular gas components. Similar to the mass outflow rate profiles, the kinetic power decreases with distance from the nucleus, as expected for centrally driven outflows expanding into the surrounding medium. The maximum kinetic power of the warm and hot phases are $\rm (1.10\pm0.03)\times10^{41}$ $\rm erg\,s^{-1}$ and {$\rm (1.22\pm0.04)\times10^{38}$ $\rm erg\,s^{-1}$}, respectively. Thus, the {total} (warm plus hot) kinetic power is {$\rm 1.10\pm0.04\times10^{41}$ $\rm erg\,s^{-1}$, corresponding to $\sim$0.015\%} of the AGN bolometric luminosity \citep[$L_{\rm bol} = \rm 7.35\times10^{44}$$\rm erg\,s^{-1}$;][]{2024MNRAS.528.1476R,2004ApJ...613..109H}.
 
\citet{2024MNRAS.528.1476R}, using optical spatially resolved observations of the central region of NGC\,3884, reported a {ionized gas} maximum mass outflow rate of $0.25\pm0.15$~M$_\odot$\,yr$^{-1}$ and a kinetic power of $(8.4\pm1.8)\times10^{40}$~erg\,s$^{-1}$, both measured at $1\farcs2$ (670~pc) from the nucleus. {Interestingly, their radial profiles show an increase in both quantities from the nucleus out to $\sim1''$, followed by a slight decrease at larger radii. This may suggest that the molecular outflow is more prominent at smaller radii, where the coupling with the molecular disk could be stronger. However, the radial profiles derived by \citet{2024MNRAS.528.1476R} cover a smaller spatial extent than those presented here, and their radial bins are smaller than the seeing of their observations. Therefore, the detailed radial behavior should be interpreted with caution.} The molecular outflows reported here exhibit approximately {a kinetic power 30\% higher than its ionized counterpart} and nearly an order of magnitude higher mass outflow rates compared to the ionized component. 

Finally, we can place constraints on the age of the outflow. Adopting an outer radius of $\sim$1.2\,kpc and assuming a constant propagation velocity of $\langle V_{out,\Delta R} \rangle \sim 500\,{\rm km\,s^{-1}}${(the average outflow velocity of the shells)}, we derive an upper limit for the onset of the outflow of $\sim$2.3\,Myr. This timescale falls withing the range expected for a typical AGN duty cycle of $\sim10^{5}-10^{6}$\,yr \citep{2015MNRAS.451.2517S,2018MNRAS.474.4740O}, further suggesting that the AGN may be heading to a fading phase. In addition, {considering that the outflow ejects mass at a constant rate we can estimate that} the central gas reservoir would last for {$M_{disk+outflows}/\dot{M}_{hot+warm}\approx$26\,Myr} if the current outflow rate is maintained {constant}. The central hole in the H$_2$ doughnut-like gas distribution may arise as a consequence of nuclear gas depletion {\citep{2021MNRAS.505L..46E,2024A&A...691A.178B}} or from H$_2$ dissociation driven by the intense radiation field {\citep{1999RvMP...71..173H}}.

\subsection{Comparison with 3C\:293}

NGC\,3884 was observed as part of the BAH project \citep{2025ApJ...982...69R}, in which three AGN host galaxies with enhanced H$_2$ emission were targeted using the MIRI and NIRSpec instruments in integral field unit mode. Detailed results on the molecular gas kinematics have already been presented for one of these galaxies, namely 3C\,293, in previous studies, which can be used as a comparison \citep{2024ApJ...974..127C,2026ApJ...997...40D}. NGC\,3884 is a radio-quiet host, in contrast to 3C\,293, which is a radio-loud source. The outflows in NGC\,3884 are likely driven by AGN radiation pressure. The higher power-law index {(see Sec. \ref{sec4.1})} may indicate that the outflows have cooled as they expanded, while the molecular disk continued to be heated by the AGN. In 3C\,293, the {outflowing} gas is kept hot by the radio jet, leading to overall higher temperatures and a lower power-law index compared to the disk component.

 Even though NGC\:3884 is one order of magnitude brighter { than 3C293}, its {mass outflow rate is} $\sim$4 times lower, and {its outflow reaches only half the distance of the one driven by the jet in 3C293}. This can be explained by the fact that mechanically-driven outflows, such as one produced by jets, typically couples more efficiently with the surrounding gas than purely radiatively-driven outflows, transferring a larger fraction of its energy and momentum to the {ISM} {\citep[e.g.][]{2020ApJ...896..114D}}. In addition, the excitation in 3C\:293 is dominated by shocks driven by the radio jets, which efficiently clear paths through the {ISM}, whereas in NGC\:3884 the excitation is likely caused by a combination of ionizing radiation and weaker shocks produced by the AGN activity.

Overall, both NGC\,3884 and 3C\,293 host AGN-driven outflows. NGC\,3884, being radio-quiet, exhibits weaker and more compact outflows likely powered by radiation pressure, whereas 3C\,293, a radio-loud source, shows mechanically-driven jets that produce stronger, more extended outflows with shock-dominated excitation. This comparison highlights how the mode of AGN feedback governs the physical properties of the surrounding gas in each galaxy.

\section{Conclusions} \label{sec:conc}

{In this work, we presented a spatially resolved analysis of the molecular and ionized gas in the inner $\sim1.2$\,kpc of NGC\,3884 using JWST/MIRI MRS observations. We mapped the gas kinematics in the mid-infrared and investigated the excitation conditions of the H$_2$ emission, with particular emphasis on the distinction between disk and outflow components. We also characterized the warm and hot molecular gas associated with the outflow and estimated its mass, mass outflow rate, and kinetic power, allowing us to investigate the role of the low-luminosity AGN in shaping the molecular gas in the circumnuclear region}. Our main results are summarized as follows.

\begin{itemize}
\item {we fitted both the ionized and molecular gas with two narrow and one broad component}. The double-peaked narrow components are consistent with gas located in the plane of the galaxy disk, as they are clearly observed along the minor axis with similar amplitudes, while their intensity ratio varies systematically along the major axis. The broad component, in contrast, is associated with an AGN-driven outflow.

\item The disk component exhibits a velocity field with a clear rotational pattern, showing a projected velocity amplitude of $\sim$150~km~s$^{-1}$ and overall low {velocity dispersion ($W_{80} \sim 200-500$~km~s$^{-1}$)}. Along the galaxy minor axis, higher turbulence is observed ($W_{80} \gtrsim 400$~km~s$^{-1}$), likely caused by the interaction of the outflow with the disk gas.

\item We detect warm molecular and ionized gas outflows extending up to $\sim$1.2~kpc from the galaxy center. The peak mass-outflow rate of the molecular component is estimated to be {$\sim 1.7~M_\odot~{\rm yr}^{-1}$} within the inner  500~pc, nearly an order of magnitude higher than that measured for the ionized gas.

\item The kinetic energy injected into the {ISM} by the molecular gas outflows corresponds to {$\sim$0.015\%} of the AGN bolometric luminosity. We estimate that the nuclear molecular gas could be depleted by the outflows within a single AGN duty cycle, assuming that the outflow rate remains constant.

\item {We find that NGC\,3884 hosts a weaker and more compact molecular outflow than the radio-loud 3C\,293, despite its higher luminosity. Although both galaxies were selected by the BAH project based on their enhanced H$_2$ emission, their different radio properties likely reflect distinct feedback mechanisms: a predominantly radiatively driven outflow in NGC\,3884 and a jet-driven, shock-dominated outflow in 3C\,293.}

\end{itemize}

{This work adds results to the still limited observational characterization of the warm molecular phase in the context of AGN feedback. The detection and characterization of a warm molecular outflow in NGC\,3884 highlight the importance of studying feedback across multiple gas phases, as its impact may be missed when considering only the ionized or cold molecular components. Our results also support the use of enhanced H$_2$/PAH ratios as a potential tracer of shocks associated with AGN-driven outflows, as proposed by \cite{2020MNRAS.491.1518R}. A natural extension of this work, already planned for future studies, is to characterize the cold molecular gas traced by CO, providing a more complete view of the multiphase gas emission structure and kinematics and its impact on the circumnuclear molecular reservoir.}

\begin{acknowledgments}
{We thank the anonymous referee for the constructive and insightful comments, which helped improve the clarity and quality of this manuscript.} JHCS  acknowledges the support from the Coordenação de Aperfeiçoamento de Pessoal de Nível Superior (CAPES; Finance code 001) and  Conselho Nacional de Desenvolvimento Científico e Tecnológico (CNPq;  Project 441722/2023-7). RAR thanks CNPq (Projects 303450/2022-3, 441722/2023-7 and 403398/2023-1), CAPES (Project 88887.894973/2023-00), and Fundação de Amparo à Pesquisa do Estado do Rio Grande do Sul (FAPERGS; Project 25/2551-0002765-9) for support. 
RR acknowledges support from  Conselho Nacional de Desenvolvimento Cient\'{i}fico e Tecnol\'ogico  ( CNPq, Proj. CNPq-445231/2024-6,311223/2020-6, 404238/2021-1, and 310413/2025-7), Funda\c{c}\~ao de amparo \`{a} pesquisa do Rio Grande do Sul (FAPERGS, Proj. 19/1750-2 and 24/2551-0001282-6) and Coordena\c{c}\~ao de Aperfei\c{c}oamento de Pessoal de N\'{i}vel Superior (CAPES, 88881.109987/2025-01). MSZM acknowledges financial support from Coordena\c c\~ao de Aperfei\c coamento de Pessoal de N\'ivel Superior (CAPES; Finance Code 001).

We thank the Mikulski Archive for Space Telescopes (MAST) staff for the data management and support.

\end{acknowledgments}

 \section*{Data Availability}
The data analyzed in this work were obtained as part of the JWST Cycle~1 program (ID~1928). The full dataset is available through the Mikulski Archive for Space Telescopes (MAST) at the Space Telescope Science Institute and can be accessed via \dataset[DOI: 10.17909/ky9r-gz73]{https://doi.org/10.17909/ky9r-gz73}.

\facilities{JWST (MIRI MRS), MAST, NED.}

\software{{\sc astropy} \citep{2022ApJ...935..167A},
{\sc ifscube} \citep{2021MNRAS.507...74R},
{\em JWST} Science Calibration pipeline \citep{2024zndo..12692459B}, 
Matplotlib \citep{hunter07}, 
QFitsView \citep{ott12},
SciPy \citep{SciPy}.}

\appendix

\section{Spatial Distribution of the Narrow Kinematic Components}\label{sec:appendixA}

{To characterize the spatial variation in the relative strength of the two narrow Gaussian components, we calculate the normalized flux difference,
$(F_{\rm blue}-F_{\rm red})/(F_{\rm blue}+F_{\rm red})$, where $F_{\rm blue}$ and $F_{\rm red}$ are the fluxes of the blue- and red-shifted narrow components, respectively. This quantity provides a direct measure of the relative contribution of each component and is shown in the left panel of Fig.~\ref{fig:ratio}. {The left panel in Fig.~\ref{fig:ratio} shows that each of the narrow gaussian components dominates one of the sides of the gas disk, while both contributes almost equally for the line flux around the minor axis}. The right panel shows the centroid velocity obtained by combining {the two narrow components as a weighted flux velocity}, in units of km\,s$^{-1}$. }

\begin{figure}[H]
    \centering
    \includegraphics[width=0.95\linewidth]{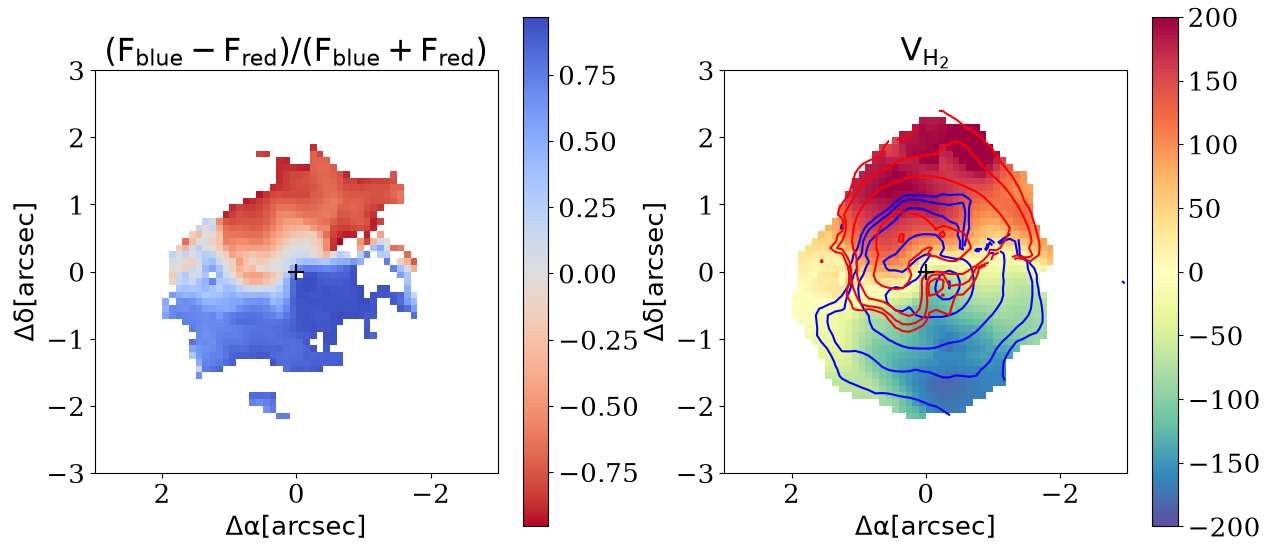}
    \caption{Spatial distribution of the two narrow Gaussian components. The left panel shows the normalized flux difference, $(F_{\rm blue}-F_{\rm red})/(F_{\rm blue}+F_{\rm red})$, between the blue- and red-shifted narrow components. The right panel shows the centroid velocity obtained by combining the two narrow components, in km\,s$^{-1}$, with contours indicating their respective flux distributions. The black cross marks the position of the emission peak in both panels.}
    \label{fig:ratio}
\end{figure}

\section{Parameter Distributions of the Narrow Components}\label{sec:appendixB}

We show in Figure~\ref{fig:dists} the parameter distributions obtained from the fitting of the two narrow Gaussian components of the H$_2$\,0--0\,S(5) emission line, which traces the emission from the disk. The blue distribution corresponds to the parameters obtained for the blueshifted component, while the red distribution represents the redshifted component. 

\begin{figure}[h]
    \centering
    \includegraphics[width=0.95\linewidth]{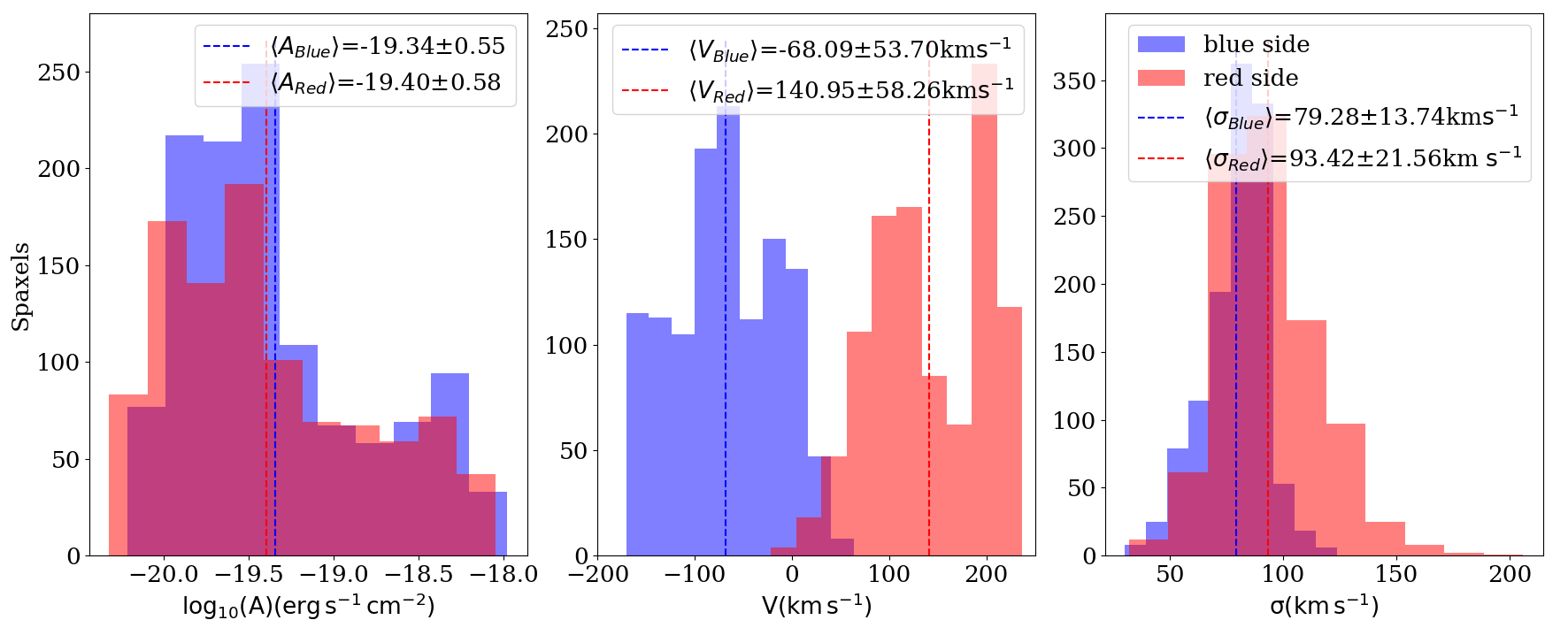}
    \caption{Parameter distributions obtained from the disk fit of the H$_{2}$\,0--0\,S(5) emission line. The different colors represent the two Gaussian components used to model the disk emission. The dashed lines indicate the mean values of the corresponding parameters: amplitude (left panel), centroid velocity (central panel), and velocity dispersion (right panel).}
    \label{fig:dists}
\end{figure}

\section{Other H$_2$ emission lines}\label{sec:appendix}

Figure~\ref{fig:appendix_maps} presents the flux distributions and kinematic maps for the $\rm H_2\,0$--$0\,S(1)$, S(3), and S(6) emission lines, showing both the disk and outflow components. These maps display patterns similar to those of the $\rm H_2\,0$--$0\,S(5)$ line shown in Fig.~\ref{fig:maps}.

\begin{figure*}[t]
    \centering
    \includegraphics[width=0.7\linewidth]{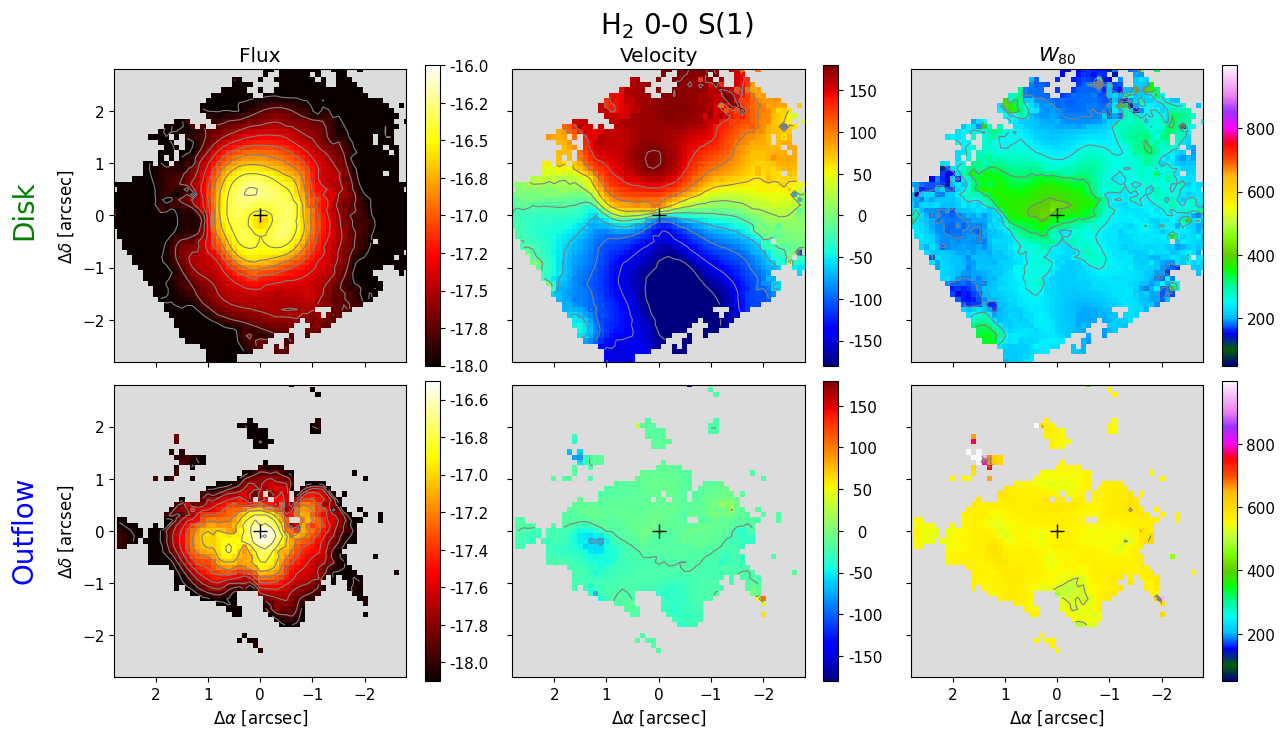}
    \includegraphics[width=0.7\linewidth]{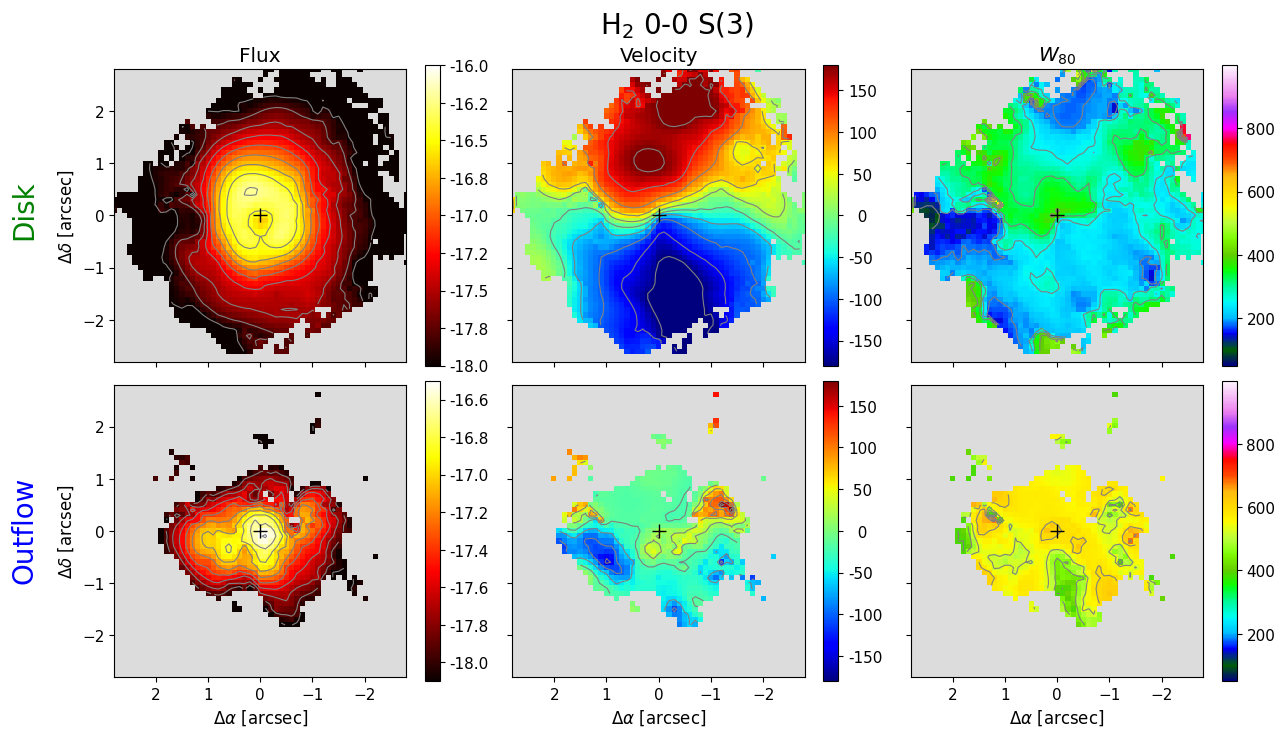}
    \includegraphics[width=0.70\linewidth]{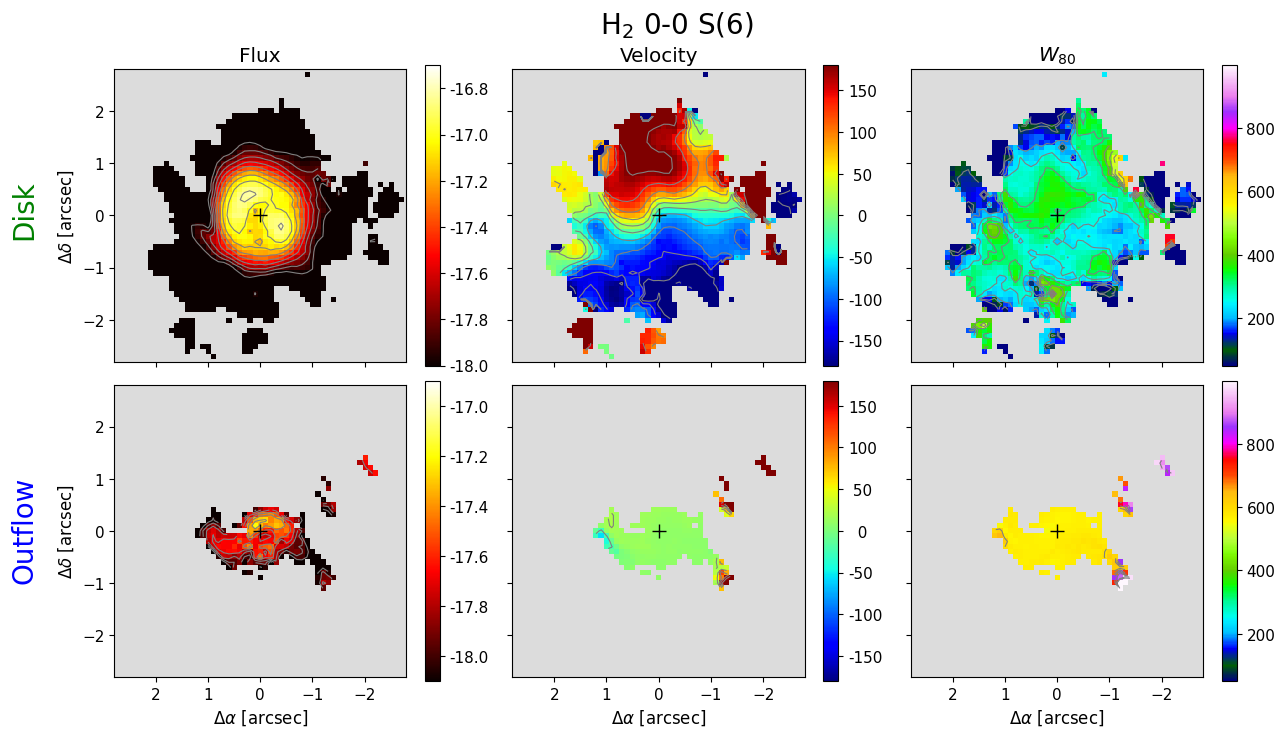}
    \caption{Same as Fig.~\ref{fig:maps}, but for the $\rm H_2\,0$--$0\,S(1)$ (top), S(3) (middle), and S(6) (bottom) emission lines. Gray regions indicate locations with $A/N < 5$ for the S(1) and S(3) lines, while regions with $A/N < 3$ are marked for the fainter S(6) emission line. }
    \label{fig:appendix_maps}
\end{figure*}

\clearpage
\bibliography{sample701}{}
\bibliographystyle{aasjournalv7}

\end{document}